\documentclass[]{aastex631}

\DeclareUnicodeCharacter{2212}{-}
\shorttitle{Rotation of a Stealth CME on 2012 October 5  Observed in the Inner Heliosphere }
\shortauthors{Kumar et al.}
\graphicspath{{./}{figures/}}

\begin{document}
\title{Rotation of a Stealth CME on 2012 October 5 Observed in the Inner Heliosphere} 

\author[0000-0002-3902-5526]{Sandeep Kumar}
\affiliation{Udaipur Solar Observatory, Physical Research Laboratory, Udaipur, 313001, India}

\author[0000-0002-6055-1192]{Dinesha V. Hegde}
\affiliation{Department of Space Science, The University of Alabama in Huntsville, AL
35805, USA}
\affiliation{Center for Space Plasma and Aeronomic Research, The University of Alabama in Huntsville, AL 35805, USA}

\author[0000-0002-0452-5838]{Nandita Srivastava}
\affiliation{Udaipur Solar Observatory, Physical Research Laboratory, Udaipur, 313001, India}

\author[0000-0002-6409-2392]{Nikolai V. Pogorelov}
\affiliation{Department of Space Science, The University of Alabama in Huntsville, AL
35805, USA}
\affiliation{Center for Space Plasma and Aeronomic Research, The University of Alabama in Huntsville, AL 35805, USA}

\author[0000-0001-5894-9954]{Nat Gopalswamy}
\affiliation{NASA Goddard Space Flight Center, Greenbelt, MD 20771, USA}
\author[0000-0002-6965-3785]{ Seiji Yashiro}
\affiliation{NASA Goddard Space Flight Center, Greenbelt, MD 20771, USA}

\begin{abstract}

Coronal Mass Ejections (CMEs) are subject to changes in their direction of propagation, tilt, and other properties. This is because CMEs interact with
the ambient solar wind and other large-scale magnetic field structures. In this work, we report on the observations of the 2012 October 5 stealth CME using coronagraphic and heliospheric images. We find clear evidence of a continuous  rotation of the CME, i.e., an increase in the tilt angle,  estimated using the Graduated Cylindrical Shell (GCS) reconstruction at different heliocentric distances, up to 58 $R_\odot$. We find a further increase in the tilt at L1 estimated from the  toroidal and cylindrical flux rope fitting on the in situ observations of IMF and solar wind parameters. This study highlights the importance of observations of Heliospheric Imager (HI), onboard the Solar TErrestrial RElations Observatory (STEREO). In particular, the GCS reconstruction of CMEs in HI field-of-view promises to bridge the gap between the near-Sun and in-situ observations at the  L1. The changes in the CME tilt has significant implications for the space weather impact of stealth CMEs.

\end{abstract}



\section{Introduction} 
\label{sec:intro}
Coronal Mass Ejections (CMEs) are the magnetic and kinetic storms in the heliosphere. CMEs are one of the mechanisms through which energy is released into the heliosphere. CME studies are of interest from both scientific and societal points of view. Scientifically, CMEs provide an excellent opportunity to understand the energy build-up and removal processes in the corona \citep{Low:1996}. CMEs are the primary cause of severe intense geomagnetic storms and solar energetic particles (SEPs). Geomagnetic storms can result in the enhancement of radiation belt particles that affect satellites in various ways. SEPs can also directly damage spacecraft. Thus, understanding CMEs and predicting their trajectory, time of arrival, and their impact on Earth are important for space weather forecasts. 

Propagation of CMEs through the interplanetary medium is quite complex, involving expansion, rotation, and deflection due to various processes. \cite{Macqueen:1986} reported that CMEs generally tend to deflect towards the equator from their respective source region near the solar minimum. Their study was based on 29 events observed by the Skylab coronagraph, which suggested that the overall bipolar magnetic field of the Sun during solar minimum may be responsible for this deflection. Deflections of a larger magnitude ($\approx 30^\circ$) were recognized by comparing the position angles of the CME leading edge and the associated prominence core \citep{gop:2000,natgop:2000,fil:2000}. \cite{Cremades:2004} suggested deflections of CMEs by $20^\circ$ towards lower latitudes due to fast solar wind flow from the polar coronal holes during solar minimum, i.e., 1996-1998. Such deflections can explain why CMEs originating at higher latitudes (up to $50^\circ$) were observed as magnetic clouds at 1 AU \citep{gop:2008} during the rising phase of solar cycle 23. In contrast, no such deflections were observed for the events near the solar maximum, i.e., 1999-2002. Magnetic fields in low-latitude coronal holes that occur predominantly in the declining phase of the solar cycle deflect CMEs away or toward the Sun-Earth line depending on their relative location \citep{gop:2008,Gop:2009,gop:2010}. 

\cite{Shen:2011} studied the kinematic evolution of a slow CME observed on 2007 October 8. They found that the gradient of radial field energy in the $\theta-\phi$ (heliographic latitude and longitude) sphere (constant R), pushed this CME in the regions of lower energy density, i.e., the region of the Heliospheric Current Sheet (HCS). After its alignment with the HCS, the CME propagated almost radially. \cite{gui:2011} studied the propagation of 10 CMEs observed during 2007-2008 by applying the method of \cite{Shen:2011} and found that the deflection in 8 of the CMEs are consistent in  strength and direction with the gradient of the magnetic energy density. An East-West asymmetry in the distribution of the sunspots related to geomagnetic storms has been known for a long time \citep{newton:1943} and has been confirmed using modern data by \citet{Wang:2002}. The East-West asymmetry of the Earth encountering halo CMEs has been found to be dependent upon the speed of the CMEs with respect to ambient solar wind speed. The source distribution of fast CMEs was found to be shifted toward the west limb of the Sun. Whereas source distribution of slow CMEs was shifted toward the East limb. This asymmetry was related to the longitudinal deflection of the CMEs by the solar wind in interplanetary space \citep{gosling:1987, Wang:2004}. 
 
A CME on 2008 September 12 was studied by \cite{Wang:2014} confirming the deflection of the CME not only in the corona but also in the heliosphere.  \citet{kay:2017} studied the deflection and rotation of CMEs which erupted from active region 11158 between February 13 and 16, 2011. They used the GCS and Forecasting CME's Altered Trajectory (ForeCAT) models to simulate the nonradial dynamics of CMEs driven by the magnetic forces and  reported rotation of the CMEs in the range of $5^\circ$ to $50^\circ$ in both clockwise and anticlockwise directions.  \citet{gop:2022} also analyzed the solar and interplanetary causes of the third largest geomagnetic storm of the solar cycle 24. They reported a prolonged acceleration of the associated CME which occurred on 2020 August 18, due to continued magnetic reconnection at the source region. They found multiple coronal holes and high-speed streams near the filament channel. The combined effect of these processes produced a complex rotation of the CME in the corona and interplanetary medium, resulting in a high-inclination magnetic cloud (MC) with a south-pointing axial magnetic field.

When a CME is shock-driving (super-Alfv\'enic), it is surrounded by a shock sheath discernible in coronagraph images, first demonstrated by \cite{sheeley:2000} using the images from Large Angle and Spectrometric Coronagraphs (LASCO, \cite{lasco:1995}) on board the Solar and Heliospheric Observatory (SOHO). As a single view is insufficient to estimate the three-dimensional (3D) shapes of CMEs, a full 3D reconstruction based on geometric modeling \citep{Thernisien_2006} became possible after the advent of the Solar TErrestrial RElations Observatory (STEREO) mission \citep{kaiser:2008}.

Researchers generally use the images of COR1 \& COR2 coronagraphs of the Sun-Earth Connection Coronal and Heliospheric Investigation (SECCHI) \citep{secchi:2008} on board STEREO and LASCO/C2 \& C3 onboard SOHO for estimating the 3D structure, direction of propagation, and kinematics of the CMEs. \cite{Wageesh:2014} utilized $J$-maps made from Heliospheric Imager (HI1 and HI2) observations on board STEREO to study interacting CMEs and the change in their dynamics. Their study showed that interaction between CMEs in the heliosphere could sometimes alter the kinematics of the CMEs.
We present a detailed analysis of the CME observed on 2012 October 5. This CME was interesting because it led to a strong geomagnetic storm during 2012 October 8-9. The Dst profile showed a two-step decrease, firstly up to $-95$ nT in the sheath region and secondly up to $-105$ nT in the MC region. Our motivation to investigate this event is stipulated by a significant mismatch between the tilt of the CME estimated from the polarity inversion line on the solar surface and those of the MC, observed at L1 \citep{Maru:2017,Martini:2022}. We demonstrate that this discrepancy can be explained by a continuous rotation of the CME flux rope between the Sun and Earth, resulting in a high inclination magnetic cloud at L1.\\
 
\section{OBSERVATIONS AND ANALYSIS}

The near-Sun observations of the 2012 October 5 CME have been studied in detail by \citet{nitta:2017}. Using the percent difference technique, they identified the source region of the CME and the tilt of the initial Polarity Inversion Line (PIL) on the Sun. However, near the Sun, this CME was deemed to be a stealth CME making it difficult to capture the signatures of the eruption on the disk. When it appeared in the LASCO/C2 field of view (FOV) it was observed as a partial halo CME. The CME was initially seen to accelerate slowly. \cite{nitta:2017} also proposed three possible source regions for the CME. These regions were not directly associated with CME, but might have contributed to the destabilization of the flux rope. Based on their analysis, they concluded that the average location of the CME source region was S25W13.

A prolonged southward component of the IMF is one of the key factors responsible for most of the intense geomagnetic storms. This depends upon the orientation of the MC associated with a CME. The tilt angle of the CME flux rope observed at 1 AU is generally assumed to remain the same as the one estimated from the PIL at the Sun, if the flux rope is unaffected by the ambient structures due to interactions \citep{gop:2022}. Therefore, it is important to examine the continuous evolution of the flux rope parameters, in particular the tilt as the CME propagates from the Sun to the Earth. For this purpose, we considered the following approach:
\begin{enumerate}
    
    \item In the outer corona from COR1 FOV, up to the time when the CME is visible in the HI1 images, we implemented the GCS reconstruction ( R$\leq$ 60 $R_\odot$), using a newly developed python module. The GCS reconstruction is a forward modeling technique to describe the self-similar expansion of a CME \citep{Thernisien_2006}. GCS  reconstruction model fit a flux rope to a CME defined by six parameters:
 latitude ($\theta$) and longitude ($\phi$) of the axis of the flux rope representing the direction of propagation of the CME; the half angle ($\alpha$) representing the width of the CME between the conical legs;  the tilt ($\gamma$) representing the orientation of the flux rope with respect to the ecliptic plane; the height ($h$) of the flux rope from the sun and the aspect ratio ($\kappa$) shows the bulkiness of the flux rope around the GCS axis.  
    
    \item At L1, we fitted the in-situ magnetic field observations with the  cylinder and torus models developed by \cite{Maru:2007}.
\end{enumerate}

\begin{figure}[!htt]
    \includegraphics[width=18.4cm]{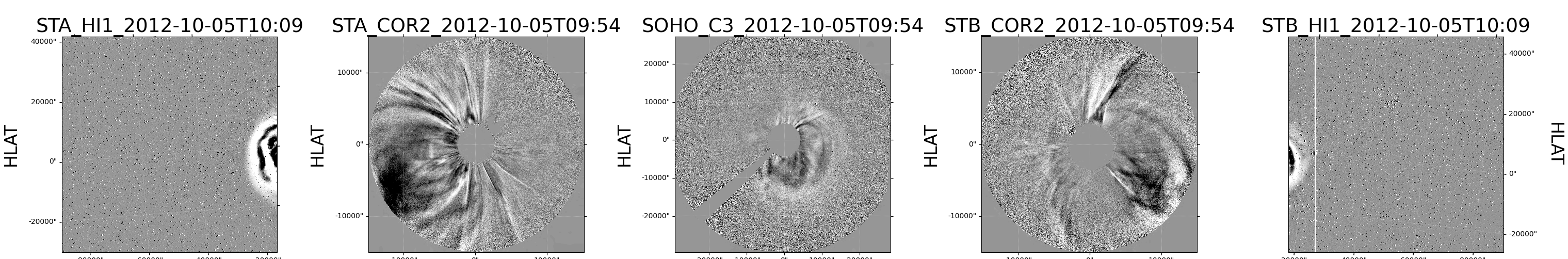} 
    \includegraphics[width=18.4cm]{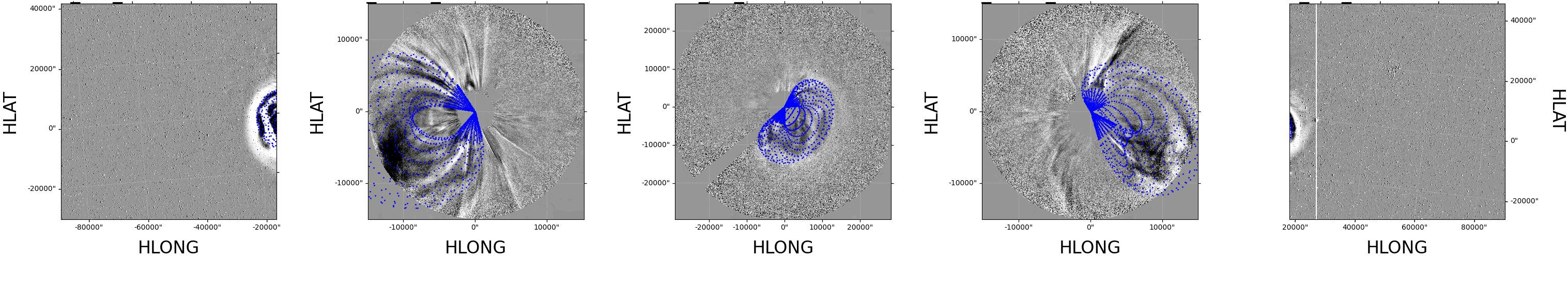}
    \caption{ Snapshots of the 2012 October 5 CME in FOV of coronagraphs and heliospheric imager (upper panel) with the corresponding GCS fits (lower panel). STEREO/A (STA) HI1, COR2, respectively are shown on the left, and STEREO/B (STB) COR2, HI1 on the right and SOHO/LASCO/C3 in the middle. HI1 images are obtained at 10:09 UT and COR2 and C3 images at 9:54 UT. Here HLAT and HLONG represent heliographic latitude and longitude, respectively.} 
    \label{fig:GCS_all}
\end{figure}

Figure~\ref{fig:GCS_all} shows images of the 2012 October 5 CME at 9:54 UT with the GCS flux rope superposed. We performed the GCS fittings starting from 07:24 UT (last frame of COR2) to 20:09 UT (last HI1 frame with clear CME signature), i.e., for approximately 13 hours.

The existing open-source Python module for GCS reconstruction uses images only up to COR2 and C3 FOV \citep{johan:2021}. We enhanced this existing python module to incorporate HI1 level2 images\footnote{\url{https://stereo-ssc.nascom.nasa.gov/data/ins_data/secchi/secchi_hi/L2_11_25/}}, and were able to continuously track the CME structure up to $\approx$ 58 $R_\odot$ (Figure:~\ref{fig:GCS_all}).%
 
In order to understand the overall evolution of this CME in the heliosphere, we compared the near-Sun tilt of the flux rope with the tilt of the CME at L1 by fitting the flux rope to the observed Interplanetary Magnetic Field (IMF) vectors recorded by the ACE spacecraft \citep{ace:1998}.

 The direction of propagation, (defined by the latitude ($\theta$) and longitude ($\phi$) of the midpoint of the axis shown) of the 2012 October 5, CME is along the local heliospheric current sheet as shown in Fig:~\ref{Fig:GCS}.  Therefore, we expect that the first two parameters of the GCS model fit, latitude ($\theta$) and  longitude ($\phi$), do not change as the CME propagates through the heliosphere, an agreement with  \cite{Shen:2011}. This will be further discussed in the next section.

Further, we estimated the heliocentric distance where the CME attains a constant width. This height is a proxy for the height beyond which CME width is constant, and therefore, flux rope can be considered as stable \citep{dagnew2022effect}. The CME was observed as a limb CME from STEREO A (the angle between the STEREO-A and CME was about $94.9^\circ$). This offered a good opportunity to study the evolution of the width of CME and also to estimate the heliocentric height at which the flux rope attained maturity.\\  
To understand the ambient magnetic field and solar wind background of the CME, we also used $pfsspy$ \citep{sby2020} for Potential Field Source Surface (PFSS) extrapolation of the photospheric magnetic field of the Sun up to $2.5R_\odot$, it also gives us  an idea of the overall structure of the magnetic field even above $2.5 R_\odot$. We also used  WSA\citep{wsa_2003,Arge:2000} code for solar wind velocity background at $21R_\odot$ from NASA/CCMC \footnote{\url{https://ccmc.gsfc.nasa.gov/results/index.php}}.

 \section{Results and Discussion}
\label{ch:2}
\subsection{CME orientation near the Sun}
\begin{figure}[!ht]
    \begin{center}
    \includegraphics[width=8.5cm]{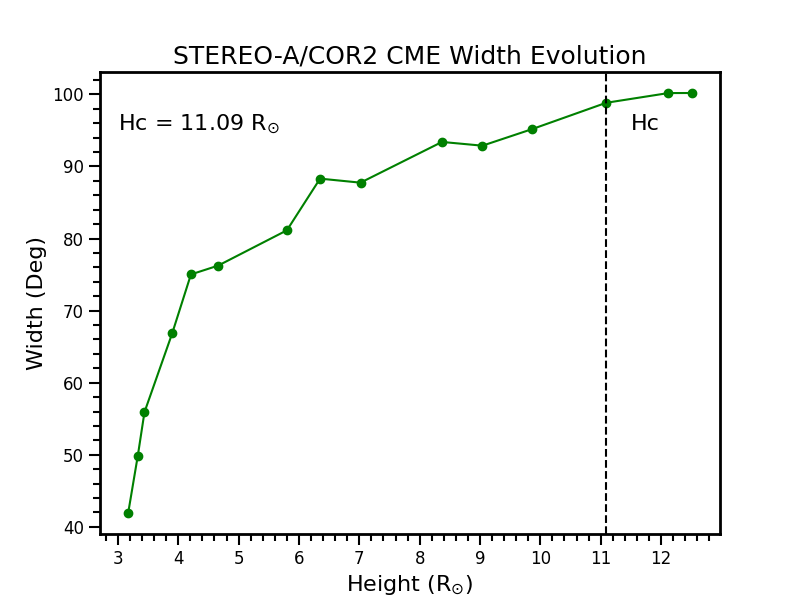} %
    \includegraphics[width=15cm]{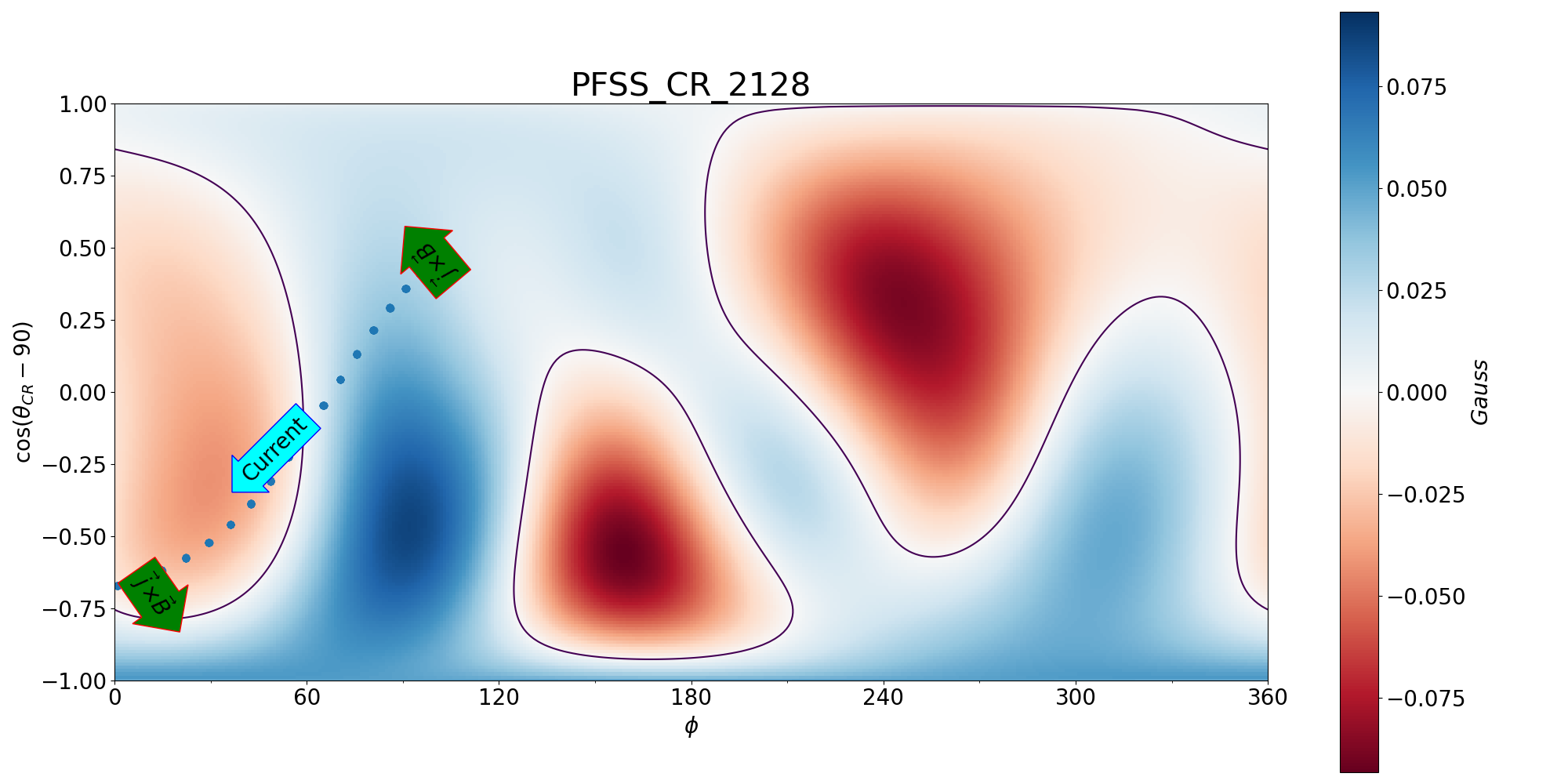}
    \caption{Top panel shows the angular width of 2012 October 5, CME in STEREO-A plotted with respect to height; here, $Hc$ represents the transition height. Bottom panel shows GCS axis (blue dots) representing current along the axis of CME front projected on the PFSS extrapolated magnetic field at $2.5R_\odot$. The cyan color arrow shows the direction of the toroidal current in the CME derived from the flux rope type in \cite{Maru:2017}, \cite{Palemrio:2018}, and \cite{Martini:2022}. The green arrows show the direction of the force on the CME.}
    \label{Fig:GCS}
    \end{center}
\end{figure}

As mentioned in the Introduction Section, it is well known that CMEs can undergo deflection due to ambient magnetic pressure that leads to  changes in the direction of their propagation, i.e., changes in latitude ($\theta$) and longitude ($\phi$) \citep{Shen:2011}, in the lower corona (COR2 FOV). Moreover, continued magnetic reconnection can lead to an increase in the width and magnetic flux of CME, i.e., an increase in half angle and kappa. These simultaneous changes can also lead to errors in the flux rope fitting of the CME flux using GCS model in the COR2 FOV, particularly in the shape and orientation of the CME flux rope.
Therefore, to clearly capture the heliospheric evolution of the CME flux rope, we start the GCS fitting from the time frame which represents a mature CME flux rope. In order to estimate the corresponding time/height of the CME, we calculate the angular width between the CME legs and the height of the leading edge/tip of the CME from the STEREO-A/COR2 images using the CDAW online measurement tool \footnote{\url{https://cdaw.gsfc.nasa.gov/movie/make_htmem_js.php?step=1&img1=sta_cor2rd&stime=20121005_0000&etime=20121005_1000}}. As mentioned earlier this was a limb event in the STEREO-A view, we can neglect the projection effects and consider the plane of sky measurements as the true measurement. The top panel of Figure~\ref{Fig:GCS} shows the plot of angular width as a function of heliocentric distance (CME Height). Here the height of the CME corresponding to the matured flux rope is referred to as the transition height (Hc) as defined by \citep{dagnew2022effect}. Figure~\ref{Fig:GCS}, shows that $Hc$ is $\approx 11 R_\odot$. Therefore, we start the GCS fitting of the stable CME after its leading edge crosses $Hc$. We choose $Hc \sim 15 R_\odot$ and start the GCS  model fitting at 7:24 UT.

 At 7:24 UT the values of best-fit parameters of GCS reconstruction, i.e., half angle, kappa, latitude, and longitude, tilt are $57^\circ$, 0.38, $-14^\circ$, $14^\circ$ and $43^\circ$, respectively, at $15.8 R_\odot$. The latitude and longitude of the GCS fit are close to the average location of the source region as identified by \cite{nitta:2017}. The values of the fitted parameters at 7:24 UT obtained here agree with those reported in \cite{Martini:2022} for the same event.

In order to study the evolution of the CME in the upcoming frames and to look for possible reasons for deflection in the direction of propagation, we examined the ambient magnetic field of the Sun as done by \citep{Shen:2011}. They reported that the CMEs can be deflected in the PIL region by the magnetic pressure gradient. This deflection can result in a change in the latitude ($\theta$) and longitude ($\phi$) parameters of the GCS reconstruction in the upcoming time steps. For the ambient medium, we used the PFSS extrapolation of the Sun from the GONG synoptic map of Carrington Rotation 2128. This extrapolation is a fairly good approximation of the overall magnetic field of the Sun. Then we plotted the GCS axis of the first GCS reconstruction overlaid on the  PFSS extrapolated magnetic field for Carrington Rotation 2128 (CR2128) as shown in Figure~\ref{Fig:GCS} (bottom panel), which shows that the direction of propagation (approximate midpoint of the axis) of the CME is in the region of the local HCS/PIL (purple line). Therefore, we do not expect any force that can significantly change the direction of propagation of the CME further \citep{Shen:2011}. Based on the above two analysis, we expected the kappa, half angle, latitude ($\theta$), and longitude ($\phi$) to remain constant above the transition height, which implies a consistent direction of propagation. We found that after the first fitting at 7:24 UT, good quality fits were obtained by changing the tilt in the upcoming time frames.

Consequently, we followed the evolution of the tilt parameter of the reconstructed CME flux rope. We considered a two-hour interval between each GCS fit to identify significant changes in the tilt parameter. We found a gradual increase in the tilt which indicates CME rotation in the heliosphere from 15 to 58 $R_\odot$. Beyond 58 $R_\odot$ it was not possible to visualize the CME clearly in the HI1 FOV therefore, GCS reconstruction could not be implemented in the HI2 FOV. 

\begin{figure}
    \centering
    \includegraphics{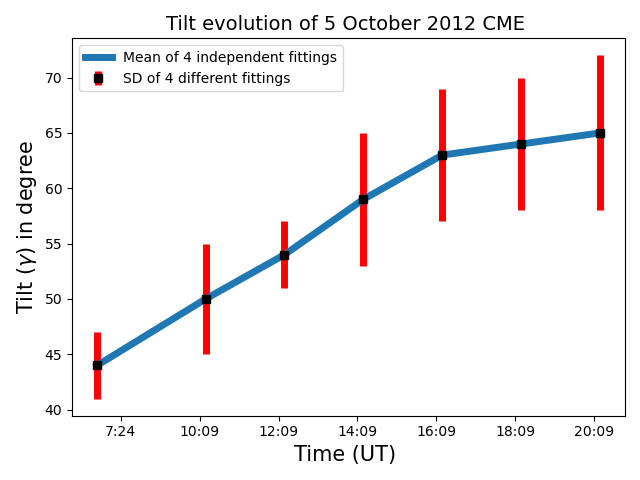}
    \caption{Evolution of tilt with respect to time of GCS fitting of the CME of 2012 October 5 in COR2 last frame and in the HI FOV. The error bars are incorporated based on four independent fittings.}
    \label{fig:tilt_err}
\end{figure}

We also examined the effect of change in the values of other GCS parameters on the values of the tilt. This was done by four independent GCS  fittings at all time steps. The independent fittings also demonstrate that the latitude and longitude are the two parameters that vary the least, i.e., not more than $3^\circ$.  Therefore,  we can safely consider the CME propagation direction to be constant.
However, to examine a different scenario, in two of these fittings, we changed  all the parameters, which showed an increase in tilt by approx $14^\circ$ in HI1 FOV. In the other two independent fittings, the half angle and kappa were assumed to be constant after the flux rope attained maturity, which showed a larger increase in tilt values, i.e., mean $\approx$ $28^\circ$.
These independent fittings suggest that a small increase in the tilt of the flux rope could result in a corresponding increase either in kappa or half-angle, as compared to the constant kappa and half-angle method of fitting.
Therefore, an assumption of constant half angle and kappa can alter the tilt values. However, it is important to mention that in two of our fittings, we consider the values of half angle and kappa  to be constant based on the above-mentioned analysis. 
All four independent GCS  fittings clearly show the increasing tilt of the flux rope, i.e., the rotation of CME in the HI FOV. The overall trend of the increase in tilt is clearly noticeable in the attached animation (in the supplementary material of the manuscript and on the Google Drive link\footnote{\url{https://drive.google.com/file/d/1OOaA5B549q6ON3UvgMxZ2EHR9vXnZzYe/view?usp=sharing}}) made from one of our independent fittings, showing $\approx$ $30^\circ$ of gradual increase in tilt. This animation also clearly demonstrates the rotation of the fitted CME flux rope (Please note that we changed the size and representations of labels in Fig:\ref{fig:GCS_all} for a better representation, as compared to the labels in animation).
We have plotted the mean values of the CME flux rope tilt obtained from four independent fittings and the standard deviation (SD) as the error in Figure~\ref {fig:tilt_err}. This shows a  $21^\circ$ increase in the mean value of the tilt from $44^\circ\pm 3^\circ$ to $65^\circ \pm 7^\circ$ from 7:24 UT to 20:09 UT (approximately  height ranging from $15\pm 1 R_\odot$ to $58 \pm 1R_\odot$). We also found as the structure evolved in the heliosphere became fainter, errors in the tilt increased as shown in the Fig:~\ref{fig:tilt_err}.

\citet{Martini:2022} used the  coronagraphic observations from LASCO/C2 FOV, COR2 FOV and from LASCO/C3 FOV. Using the GCS method in COR2 FOV and the elliptical fitting method in C2 and C3 FOV, the increase in tilt reported by them agrees with our findings. Our approach incorporates continuous tracking of the CME in HI FOV and also at L1.  This analysis provides clear evidence of the continuous rotation of the CME  throughout the heliosphere.

\subsection{CME orientation at L1}
\begin{figure}
    \begin{center}
     \includegraphics[width=19.5cm]{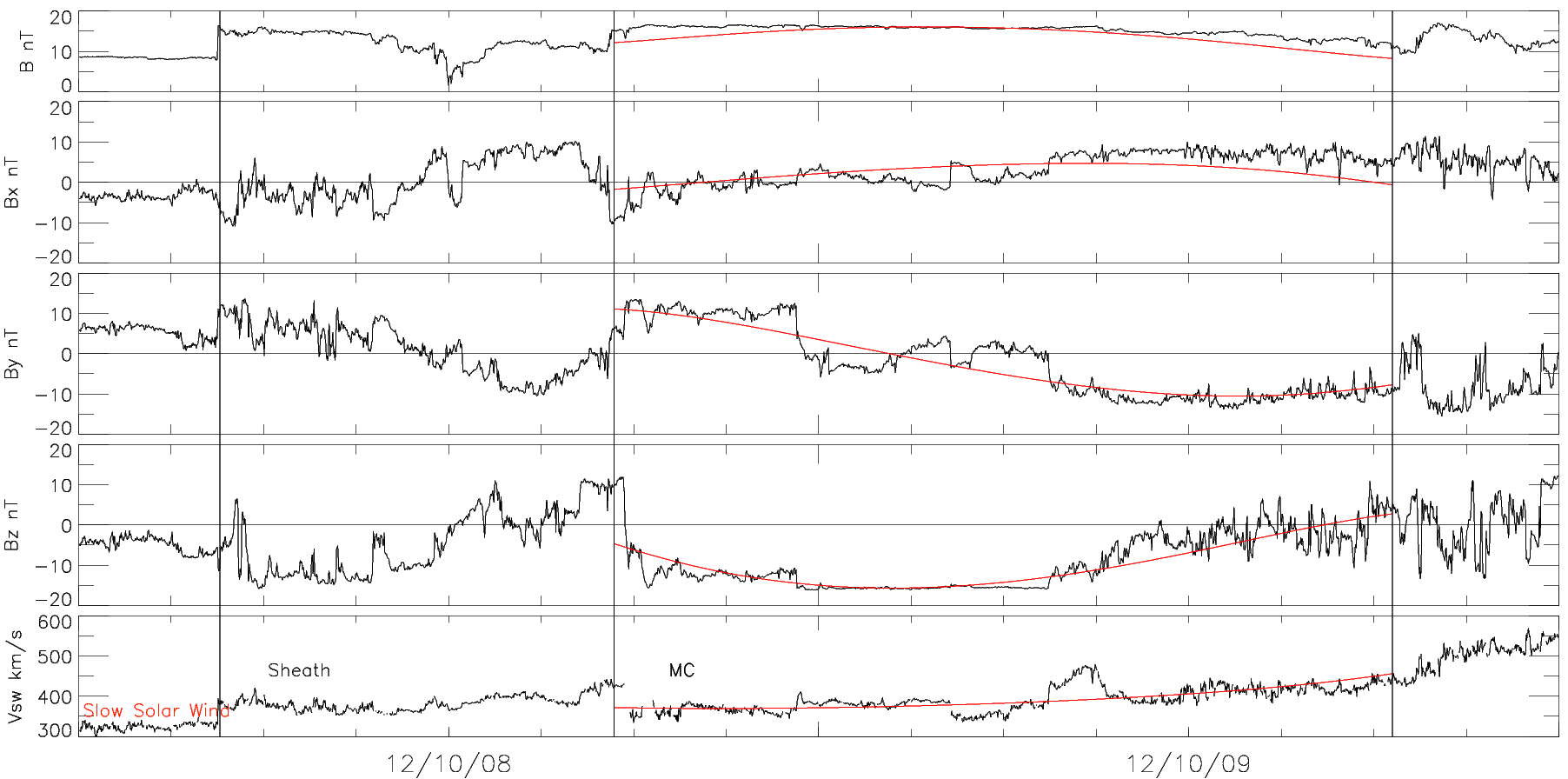} %
    \includegraphics[width=10cm]{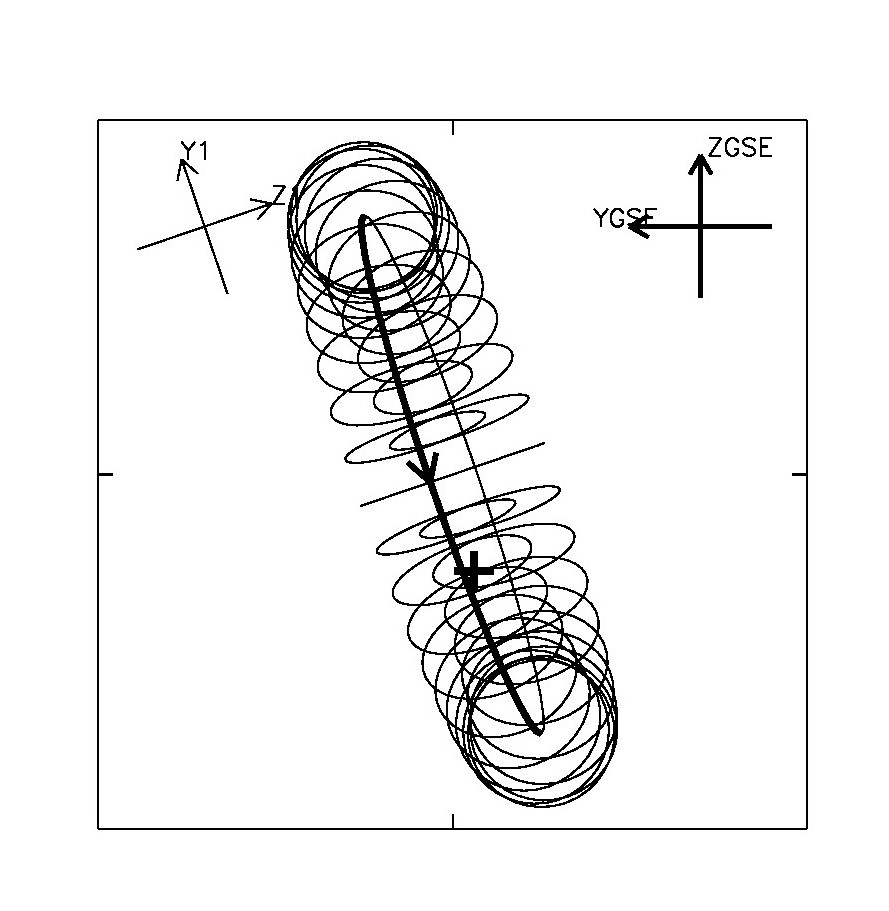} %
    \caption{Top panel shows the Marubashi fitting of Toroidal model ($Torus_{M03}$) with solar wind parameters in the magnetic cloud (red line) and observed solar wind parameters from ACE plotted with respect to time (YY/MM/DD), i.e., $|B|$, $B_x$, $B_y$, $B_z$, of IMF and bulk solar wind velocity in kms$^{-1}$ from top to bottom respectively (black curve). The bottom panel shows the orientation of the fitted flux rope at L1 for $Torus_{M03}$. Here $+$ sign shows the position of the spacecraft crossing the magnetic cloud. We measured the tilt angle anti-clockwise from the negative y-axis of the GSE coordinate system.}
    \label{Fig:Maru}
    \end{center}
\end{figure}

As mentioned earlier, we used the cylindrical and toroidal models of \cite{Maru:2007} to fit the flux rope to the observed IMF at L1. These two models are force-free constant-$\alpha$ models, providing  the latitude and longitude of the flux rope axis in GSE coordinates, which can be used to derive the tilt of the axis.
Figure~\ref{Fig:Maru} (top panel) shows the observed time evolution of solar wind IMF parameters  $|B|$, $B_x$, $B_y$,  and $B_z$, and bulk solar wind velocity (black curves). The plot shows the passage of the ICME from 17:22 UT  on 8 October to 18:36 UT on 9 October 2012. It includes the shock arrival marked by the leftmost vertical line, followed by the sheath region between the left and middle vertical lines. The region between the middle and rightmost vertical lines represents the MC interval. The MC exhibits rotation in the $B_y$ component. This shows a slow solar wind, turbulent sheath region, and the MC region. The sheath region and MC showed a prolonged southward IMF. The MC alone showed southward IMF for almost 25.5 hours, which caused a geomagnetic storm of Dst -105 $nT$. Based on the cylindrical flux rope fitting (tilt $\approx 114^\circ$)  and the fitting of the toroidal flux rope (tilt $\approx 108^\circ$), we estimate the tilt of the flux rope at L1 to be $\approx 110^\circ$. This high inclination flux rope is also consistent with the rotation observed in the $B_y$ and negative $B_z$ throughout the MC.\\ 
 The upper plot of Figure:~\ref{Fig:Maru} shows the Marubashi fitting of toroidal model $Torus_{M03}$ (our best fit to the observed solar wind parameters) to the MC (red line). The bottom panel of Figure:~\ref{Fig:Maru} shows the orientation of the fitted flux rope at L1 for for this model. Here the $+$ sign shows the position of the spacecraft crossing the MC. In order to compare the tilt of the flux rope obtained by GCS fit near the Sun with that estimated at L1, we measure the tilt angle at L1, in the anti-clockwise direction from the negative y-axis of the GSE coordinate system.

\begin{figure}[!ht]
    \begin{center}
    
    \includegraphics[width=15cm]{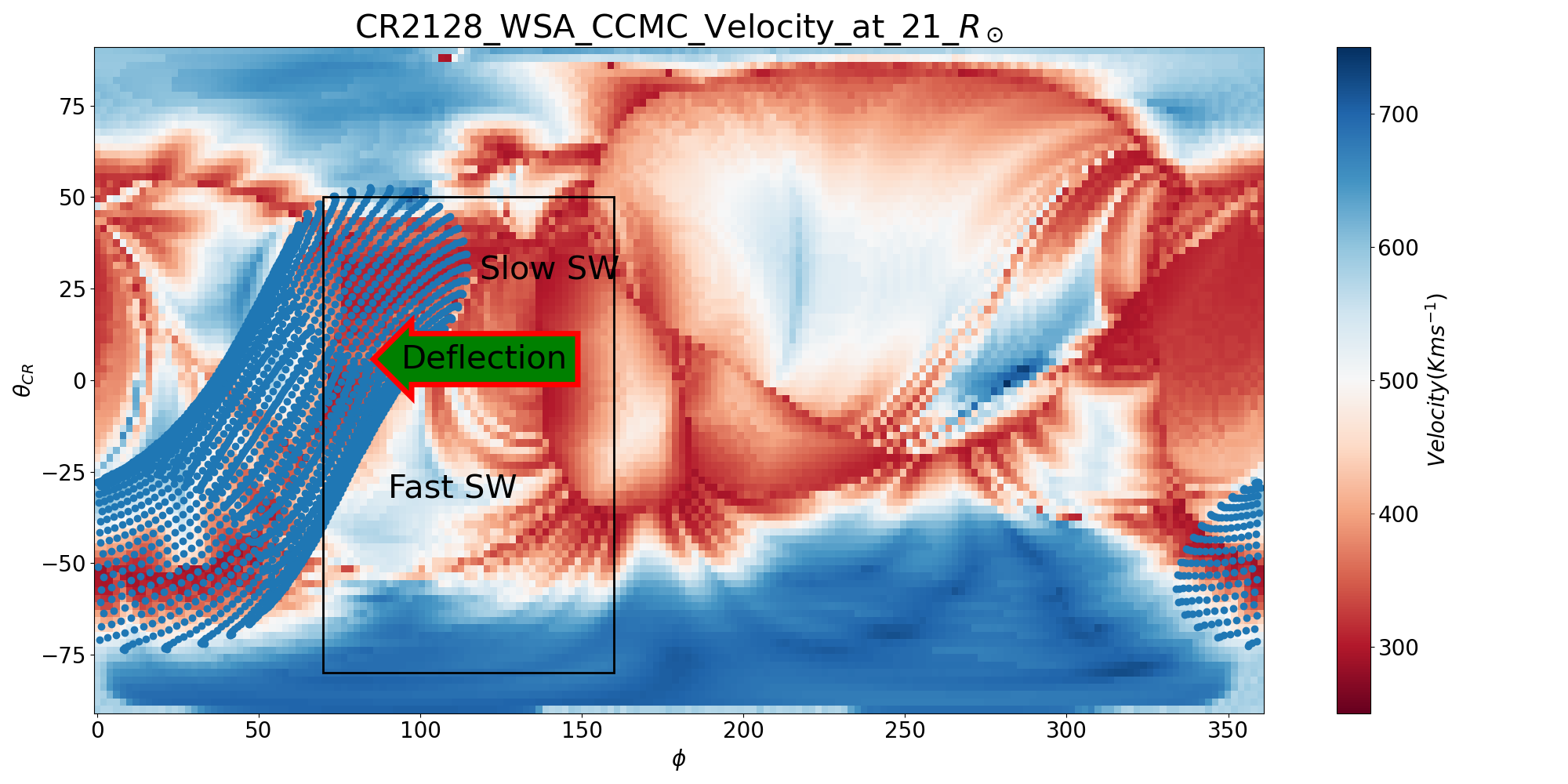}
    \caption{This figure shows the WSA velocity background of the CME at 21$R_\odot$ from NASA/CCMC. The black rectangle specifies the in-homogeneous velocity background around the CME top front. The green arrow in the lower panel shows the deflection of the upper front of the CME due to interaction with the slow solar wind in front of it.}
    \label{fig:GCS_rot}
    \end{center}
\end{figure}

An earlier study by \cite{Maru:2017} reported that the tilt of the PIL associated with this CME was around $30^\circ$. We observed a large change in orientation, i.e., in the tilt value, $\approx$ $66^\circ$($110^\circ-44^\circ$) in the anticlockwise direction of the CME from 15 $R_\odot$ to L1. The analysis further suggests that  $\approx 30\%$ of the total change in the tilt occurred below 60 $R_\odot$ and the rest in the remaining 160 $R_\odot$, thereby suggesting a continuous rotation throughout the heliosphere. Further, we note that this rotation resulted in a highly inclined flux rope at L1, leading to a prolonged ($\approx$ 25.5 hours) southward component of IMF, resulting in an intense geomagnetic storm.

The increase in tilt observed near the Sun can be justified by the force acting on the toroidal current in the CME front (derived from the type of flux rope), by the ambient radial magnetic field of the Sun. Earlier observers reported an ESW type of flux rope near the Sun which had Right-Handed (RH) chirality with axial field pointing southward \citep{Maru:2017,Palemrio:2018}. The direction of the current derived from the ESW type of flux rope is represented by the cyan color arrow in bottom panel of Figure:~\ref{Fig:GCS}. For ambient radial magnetic field, we used PFSS extrapolation of the photospheric magnetic field up to 2.5 $R_\odot$ from pfsspy python code \citep{david_stansby_2019_2566462} as shown in the bottom panel of Figure:~\ref{Fig:GCS}. This extrapolation gives an approximation of the magnetic field polarity and magnitude between 2.5 $R_\odot$ and 21 $R_\odot$. Beyond 2.5 $R_\odot$, nearly radial extrapolation of the field can be approximated from PFSS retaining polarity and PIL structure. Further, the torque due to the ambient radial magnetic field is acting on the CME front, making the CME rotate in an anti-clockwise direction (bottom Figure:~\ref{Fig:GCS}). However, this sense of rotation may hold good only up to 21 $R_\odot$ because of the dominance of the radial magnetic field within this distance \citep{Schatten:1972}.\\
 The synoptic map of the ambient solar wind for CR2128 at 21$R_\odot$, from NASA/CCMC \footnote{\url{https://ccmc.gsfc.nasa.gov/}}, is shown in the bottom panel of Figure:~\ref{fig:GCS_rot}. 
Beyond 21 $R_\odot$, the dynamics of the ambient medium are dominated by the velocity of the solar wind \citep{Riley:2011}. Therefore, we conclude that the observed rotation in the heliosphere beyond 21 $R_\odot$ may be due to the interaction of the CME flux rope with the ambient solar wind.\\
We found a non-uniform solar wind velocity environment in the leading part of this CME as shown in Figure:~\ref{fig:GCS_rot} (black rectangle). We found a slow solar wind ($\approx$ 300 km/s) flow ahead of the top part of the CME (above the green arrow, Figure:~\ref{fig:GCS_rot}). In contrast, the lower part of the CME faces an overall fast solar wind background in front of it (below the green arrow).
The slow solar wind heading the top part of the fast CME ($v \approx 600$ $km/s $) can deflect the upper part of the CME in the eastward direction due to rotation of the Sun, as shown by the green arrow. 
Moreover, this kind of deflection is not possible in the lower part of the CME, as shown in the lower half of the black rectangle, because it faces an overall fast solar wind in front of it. This rotates the CME in the same sense in the rest of the heliosphere as provided by the force due to the ambient radial magnetic field on the CME axis, below $21R_\odot$. The slow solar wind ahead of the CME front is further confirmed by the slow solar wind velocity observed in in-situ observations at L1 by ACE (top panel, Figure:~\ref{Fig:Maru}). Earlier studies have shown eastward deflections of fast CMEs propagating in the slow solar wind, these studies did not take into account the latitudinal extension of the CMEs in the heliosphere, whereas we base 
our arguments on the possibility that different parts of a CME can propagate through different environments, thereby rotating the CME as it propagates \citep{Wang:2004,Gop:2009}. 
We, therefore, believe that the interaction of the CME with the solar wind in the rest of the heliosphere led to a further increase in the tilt at L1 (Figure:~\ref{Fig:Maru}). The analysis suggests that a favorable solar wind environment can change the orientation leading to rotation of the CME, resulting in a geoeffective event. A recent study by \cite{He:2018} also suggested an increase in the geoeffectiveness of a stealth CME on 2016 October 8, because of its interaction with ambient solar wind medium. Their study showed the increased geoeffectiveness was due to the interaction of the CME with the Co-rotating Interacting Region (CIR) in the ambient solar wind. Our study further supports the role of interaction of the ambient solar wind with the stealth CME,  enhancing its geoeffectiveness at the Earth.
\section{Conclusions}
\begin{enumerate}
    \item The study of the evolution of 2012 October 5 CME in the heliosphere shows clear evidence of rotation of the CME as indicated by a continuous increase in the tilt from COR2 FOV, starting from $15R_\odot$ to HI1 FOV (up to $58R_\odot$).  Although it was not possible to track the CME in the HI FOV beyond $58R_\odot$, the flux rope fitting implementation to in-situ observations using the cylindrical and toroidal model of flux rope by \cite{Maru:2007} resulted in an increased value of the tilt at the L1 point as compared to that measured in the last fame of HI1. This suggests a continuous increase in tilt from near Sun to L1.
    \item The 2012 October 5  CME did not leave any low coronal signatures on the disk, making it difficult for forecasters to assess its impact on Earth. The CME propagated at a moderate speed of 600 $kms^{-1}$ near the Sun. However, it experienced a continuous increase in its tilt due to its propagation from the Sun to the Earth. This led to a prolonged southward component of the flux rope, which was responsible for its enhanced geoeffectiveness. Our results further highlight the challenges in space weather forecasting of such stealthy CMEs. Our study demonstrates that  it may be difficult sometimes to forecast the geoeffectiveness of the CMEs on the basis of the near-Sun observations alone. In this regard, HI observations prove to be crucial which bridge the gap between the near-Sun and near-Earth observation, thereby providing an improved understanding of CME propagation in the heliosphere. 
\end{enumerate}

\begin{acknowledgments}
We thank the reviewer for the constructive comments to improve the manuscript. We thank and acknowledge K. Marubashi for providing us the flux rope fitting code in IDL. We also acknowledge the use of ACE magnetic field and velocity data (\url{https://izw1.caltech.edu/ACE/ASC/level2/lvl2DATA_MAG.html}) and GONG program for the synoptic map data (\url{https://gong.nso.edu/data/magmap/crmap.html}.). We acknowledge the use of the GCS and $pfsspy$ code. We acknowledge the sunpy community\citep{sunpy_community2020}. We acknowledge the NASA/CCMC program for providing us the WSA velocity map at 21 $R_\odot$. Coronagraphic images are used from Helioviewer and HI1 images from the site (\url{https://stereo-ssc.nascom.nasa.gov/data/ins_data/secchi/secchi_hi/L2_11_25/}). This work was carried out under the Indo-U.S. Science and Technology Forum (IUSSTF) Virtual Network Center project (Ref. no. IUSSTF/JC-113/2019). SK acknowledges the support from IUSSTF and the University of Alabama in Huntsville (UAH) for the visit to UAH during which this work was initiated. NG is supported by NASA's STEREO project and Living With a Star program. The work of DVH was supported by NASA FINESST grant 80NSSC22K0058. NP was supported, in part, by NSF/NASA Space Weather with Quantified Uncertainty grant 2028154. 
\end{acknowledgments}

\bibliography{manuscript}{}

\begin{thebibliography}{}
\expandafter\ifx\csname natexlab\endcsname\relax\def\natexlab#1{#1}\fi
\providecommand{\url}[1]{\href{#1}{#1}}
\providecommand{\dodoi}[1]{doi:~\href{http://doi.org/#1}{\nolinkurl{#1}}}
\providecommand{\doeprint}[1]{\href{http://ascl.net/#1}{\nolinkurl{http://ascl.net/#1}}}
\providecommand{\doarXiv}[1]{\href{https://arxiv.org/abs/#1}{\nolinkurl{https://arxiv.org/abs/#1}}}

\bibitem[{{Arge} {et~al.}(2003){Arge}, {Odstrcil}, {Pizzo}, \&
  {Mayer}}]{wsa_2003}
{Arge}, C.~N., {Odstrcil}, D., {Pizzo}, V.~J., \& {Mayer}, L.~R. 2003, in
  American Institute of Physics Conference Series, Vol. 679, Solar Wind Ten,
  ed. M.~{Velli}, R.~{Bruno}, F.~{Malara}, \& B.~{Bucci}, 190--193,
  \dodoi{10.1063/1.1618574}

\bibitem[{{Arge} \& {Pizzo}(2000)}]{Arge:2000}
{Arge}, C.~N., \& {Pizzo}, V.~J. 2000, \jgr, 105, 10465,
  \dodoi{10.1029/1999JA000262}

\bibitem[{Brueckner {et~al.}(1995)Brueckner, Howard, Koomen, Korendyke,
  Michels, Moses, Socker, Dere, Lamy, Llebaria, Bout, Schwenn, Simnett,
  Bedford, \& Eyles}]{lasco:1995}
Brueckner, G., Howard, R., Koomen, M., {et~al.} 1995, Solar Physics, 162, 357,
  \dodoi{10.1007/BF00733434}

\bibitem[{{Cremades} \& {Bothmer}(2004)}]{Cremades:2004}
{Cremades}, \& {Bothmer}. 2004, A\&A, 422, 307,
  \dodoi{10.1051/0004-6361:20035776}

\bibitem[{{Dagnew} {et~al.}(2022){Dagnew}, {Gopalswamy}, {Tessema}, {Akiyama},
  \& {Yashiro}}]{dagnew2022effect}
{Dagnew}, F.~K., {Gopalswamy}, N., {Tessema}, S.~B., {Akiyama}, S., \&
  {Yashiro}, S. 2022, \apj, 936, 122, \dodoi{10.3847/1538-4357/ac8744}

\bibitem[{{Filippov} {et~al.}(2001){Filippov}, {Gopalswamy}, \&
  {Lozhechkin}}]{fil:2000}
{Filippov}, B.~P., {Gopalswamy}, N., \& {Lozhechkin}, A.~V. 2001, \solphys,
  203, 119, \dodoi{10.1023/A:1012754329767}

\bibitem[{{Gopalswamy} {et~al.}(2008){Gopalswamy}, {Akiyama}, {Yashiro},
  {Michalek}, \& {Lepping}}]{gop:2008}
{Gopalswamy}, N., {Akiyama}, S., {Yashiro}, S., {Michalek}, G., \& {Lepping},
  R.~P. 2008, Journal of Atmospheric and Solar-Terrestrial Physics, 70, 245,
  \dodoi{10.1016/j.jastp.2007.08.070}

\bibitem[{{Gopalswamy} {et~al.}(2000){Gopalswamy}, {Hanaoka}, \&
  {Hudson}}]{gop:2000}
{Gopalswamy}, N., {Hanaoka}, Y., \& {Hudson}, H.~S. 2000, Advances in Space
  Research, 25, 1851, \dodoi{10.1016/S0273-1177(99)00597-9}

\bibitem[{{Gopalswamy} {et~al.}(2010){Gopalswamy}, {M{\"a}kel{\"a}}, {Xie},
  {Akiyama}, \& {Yashiro}}]{gop:2010}
{Gopalswamy}, N., {M{\"a}kel{\"a}}, P., {Xie}, H., {Akiyama}, S., \& {Yashiro},
  S. 2010, in American Institute of Physics Conference Series, Vol. 1216,
  Twelfth International Solar Wind Conference, ed. M.~{Maksimovic},
  K.~{Issautier}, N.~{Meyer-Vernet}, M.~{Moncuquet}, \& F.~{Pantellini},
  452--458, \dodoi{10.1063/1.3395902}

\bibitem[{Gopalswamy {et~al.}(2009)Gopalswamy, Makela, Xie, \&
  Yashiro}]{Gop:2009}
Gopalswamy, N., Makela, P., Xie, H., \& Yashiro, S. 2009, Journal of
  Geophysical Research, 114, \dodoi{10.1029/2008JA013686}

\bibitem[{{Gopalswamy} \& {Thompson}(2000)}]{natgop:2000}
{Gopalswamy}, N., \& {Thompson}, B.~J. 2000, Journal of Atmospheric and
  Solar-Terrestrial Physics, 62, 1457, \dodoi{10.1016/S1364-6826(00)00079-1}

\bibitem[{Gopalswamy {et~al.}(2022)Gopalswamy, Yashiro, Akiyama, Xie, Mäkelä,
  Fok, \& Ferradas}]{gop:2022}
Gopalswamy, N., Yashiro, S., Akiyama, S., {et~al.} 2022, Journal of Geophysical
  Research: Space Physics, 127, e2022JA030404,
  \dodoi{https://doi.org/10.1029/2022JA030404}

\bibitem[{{Gosling} {et~al.}(1987){Gosling}, {Thomsen}, {Bame}, \&
  {Zwickl}}]{gosling:1987}
{Gosling}, J.~T., {Thomsen}, M.~F., {Bame}, S.~J., \& {Zwickl}, R.~D. 1987,
  \jgr, 92, 12399, \dodoi{10.1029/JA092iA11p12399}

\bibitem[{Gui {et~al.}(2011)Gui, Shen, Wang, Ye, Liu, Wang, \& Zhao}]{gui:2011}
Gui, B., Shen, C., Wang, Y., {et~al.} 2011, Solar Physics - SOL PHYS, 271,
  \dodoi{10.1007/s11207-011-9791-9}

\bibitem[{He {et~al.}(2018)He, Liu, Hu, Wang, \& Zhao}]{He:2018}
He, W., Liu, Y.~D., Hu, H., Wang, R., \& Zhao, X. 2018, The Astrophysical
  Journal, 860, 78, \dodoi{10.3847/1538-4357/aac381}

\bibitem[{{Howard} {et~al.}(2008){Howard}, {Moses}, {Vourlidas}, {Newmark},
  {Socker}, {Plunkett}, {Korendyke}, {Cook}, {Hurley}, {Davila}, {Thompson},
  {St Cyr}, {Mentzell}, {Mehalick}, {Lemen}, {Wuelser}, {Duncan}, {Tarbell},
  {Wolfson}, {Moore}, {Harrison}, {Waltham}, {Lang}, {Davis}, {Eyles},
  {Mapson-Menard}, {Simnett}, {Halain}, {Defise}, {Mazy}, {Rochus}, {Mercier},
  {Ravet}, {Delmotte}, {Auchere}, {Delaboudiniere}, {Bothmer}, {Deutsch},
  {Wang}, {Rich}, {Cooper}, {Stephens}, {Maahs}, {Baugh}, {McMullin}, \&
  {Carter}}]{secchi:2008}
{Howard}, R.~A., {Moses}, J.~D., {Vourlidas}, A., {et~al.} 2008, \ssr, 136, 67,
  \dodoi{10.1007/s11214-008-9341-4}

\bibitem[{{Kaiser} {et~al.}(2008){Kaiser}, {Kucera}, {Davila}, {St. Cyr},
  {Guhathakurta}, \& {Christian}}]{kaiser:2008}
{Kaiser}, M.~L., {Kucera}, T.~A., {Davila}, J.~M., {et~al.} 2008, \ssr, 136, 5,
  \dodoi{10.1007/s11214-007-9277-0}

\bibitem[{{Kay} {et~al.}(2017){Kay}, {Gopalswamy}, {Xie}, \&
  {Yashiro}}]{kay:2017}
{Kay}, C., {Gopalswamy}, N., {Xie}, H., \& {Yashiro}, S. 2017, \solphys, 292,
  78, \dodoi{10.1007/s11207-017-1098-z}

\bibitem[{{Low}(1996)}]{Low:1996}
{Low}, B.~C. 1996, \solphys, 167, 217, \dodoi{10.1007/BF00146338}

\bibitem[{MacQueen {et~al.}(1986)MacQueen, Hundhausen, \&
  Conover}]{Macqueen:1986}
MacQueen, R.~M., Hundhausen, A.~J., \& Conover, C.~W. 1986, Journal of
  Geophysical Research: Space Physics, 91, 31,
  \dodoi{https://doi.org/10.1029/JA091iA01p00031}

\bibitem[{{Martini{\'c}} {et~al.}(2022){Martini{\'c}}, {Dumbovi{\'c}},
  {Temmer}, {Veronig}, \& {Vr{\v{s}}nak}}]{Martini:2022}
{Martini{\'c}}, K., {Dumbovi{\'c}}, M., {Temmer}, M., {Veronig}, A., \&
  {Vr{\v{s}}nak}, B. 2022, \aap, 661, A155, \dodoi{10.1051/0004-6361/202243433}

\bibitem[{{Marubashi} {et~al.}(2017){Marubashi}, {Cho}, \&
  {Ishibashi}}]{Maru:2017}
{Marubashi}, K., {Cho}, K.~S., \& {Ishibashi}, H. 2017, \solphys, 292, 189,
  \dodoi{10.1007/s11207-017-1204-2}

\bibitem[{Marubashi \& Lepping(2007)}]{Maru:2007}
Marubashi, K., \& Lepping, R.~P. 2007, Annales Geophysicae, 25, 2453,
  \dodoi{10.5194/angeo-25-2453-2007}

\bibitem[{{Mishra} \& {Srivastava}(2014)}]{Wageesh:2014}
{Mishra}, W., \& {Srivastava}, N. 2014, \apj, 794, 64,
  \dodoi{10.1088/0004-637X/794/1/64}

\bibitem[{{Newton}(1943)}]{newton:1943}
{Newton}, H.~W. 1943, \mnras, 103, 244, \dodoi{10.1093/mnras/103.5.244}

\bibitem[{{Nitta} \& {Mulligan}(2017)}]{nitta:2017}
{Nitta}, N.~V., \& {Mulligan}, T. 2017, \solphys, 292, 125,
  \dodoi{10.1007/s11207-017-1147-7}

\bibitem[{Palmerio {et~al.}(2018)Palmerio, Kilpua, Möstl, Bothmer, James,
  Green, Isavnin, Davies, \& Harrison}]{Palemrio:2018}
Palmerio, E., Kilpua, E. K.~J., Möstl, C., {et~al.} 2018, Space Weather, 16,
  442, \dodoi{https://doi.org/10.1002/2017SW001767}

\bibitem[{{Riley} \& {Lionello}(2011)}]{Riley:2011}
{Riley}, P., \& {Lionello}, R. 2011, \solphys, 270, 575,
  \dodoi{10.1007/s11207-011-9766-x}

\bibitem[{{Schatten}(1972)}]{Schatten:1972}
{Schatten}, K.~H. 1972, {``Current Sheet Magnetic Model for the Solar Corona"}
  in {Solar Wind}, ed. C.~P. {Sonett}, P.~J. {Coleman}, \& J.~M. {Wilcox}, Vol.
  308 (Washington: Scientific and Technical Information Office, National
  Aeronautics and Space Administration), 44

\bibitem[{{Sheeley} {et~al.}(2000){Sheeley}, {Hakala}, \&
  {Wang}}]{sheeley:2000}
{Sheeley}, N.~R., {Hakala}, W.~N., \& {Wang}, Y.~M. 2000, \jgr, 105, 5081,
  \dodoi{10.1029/1999JA000338}

\bibitem[{{Shen} {et~al.}(2011){Shen}, {Wang}, {Gui}, {Ye}, \&
  {Wang}}]{Shen:2011}
{Shen}, C., {Wang}, Y., {Gui}, B., {Ye}, P., \& {Wang}, S. 2011, \solphys, 269,
  389, \dodoi{10.1007/s11207-011-9715-8}

\bibitem[{Stansby(2019)}]{david_stansby_2019_2566462}
Stansby, D. 2019, dstansby/pfsspy: pfsspy 0.1.2, 0.1.2,  Zenodo,
  \dodoi{10.5281/zenodo.2566462}

\bibitem[{Stansby {et~al.}(2020)Stansby, Yeates, \& Badman}]{sby2020}
Stansby, D., Yeates, A., \& Badman, S.~T. 2020, Journal of Open Source
  Software, 5, 2732, \dodoi{10.21105/joss.02732}

\bibitem[{{Stone} {et~al.}(1998){Stone}, {Frandsen}, {Mewaldt}, {Christian},
  {Margolies}, {Ormes}, \& {Snow}}]{ace:1998}
{Stone}, E.~C., {Frandsen}, A.~M., {Mewaldt}, R.~A., {et~al.} 1998, \ssr, 86,
  1, \dodoi{10.1023/A:1005082526237}

\bibitem[{{The SunPy Community} {et~al.}(2020){The SunPy Community}, Barnes,
  Bobra, Christe, Freij, Hayes, Ireland, Mumford, Perez-Suarez, Ryan, Shih,
  Chanda, Glogowski, Hewett, Hughitt, Hill, Hiware, Inglis, Kirk, Konge, Mason,
  Maloney, Murray, Panda, Park, Pereira, Reardon, Savage, Sipőcz, Stansby,
  Jain, Taylor, Yadav, Rajul, \& Dang}]{sunpy_community2020}
{The SunPy Community}, Barnes, W.~T., Bobra, M.~G., {et~al.} 2020, The
  Astrophysical Journal, 890, 68, \dodoi{10.3847/1538-4357/ab4f7a}

\bibitem[{Thernisien {et~al.}(2006)Thernisien, Howard, \&
  Vourlidas}]{Thernisien_2006}
Thernisien, A. F.~R., Howard, R.~A., \& Vourlidas, A. 2006, The Astrophysical
  Journal, 652, 763, \dodoi{10.1086/508254}

\bibitem[{von Forstner(2021)}]{johan:2021}
von Forstner, J. L.~F. 2021, johan12345/gcs\_python: Release 0.2.2, 0.2.2,
  Zenodo, \dodoi{10.5281/zenodo.5084818}

\bibitem[{{Wang} {et~al.}(2004){Wang}, {Shen}, {Wang}, \& {Ye}}]{Wang:2004}
{Wang}, Y., {Shen}, C., {Wang}, S., \& {Ye}, P. 2004, \solphys, 222, 329,
  \dodoi{10.1023/B:SOLA.0000043576.21942.aa}

\bibitem[{Wang {et~al.}(2014)Wang, Wang, Shen, Shen, \& Lugaz}]{Wang:2014}
Wang, Y., Wang, B., Shen, C., Shen, F., \& Lugaz, N. 2014, Journal of
  Geophysical Research: Space Physics, 119, 5117,
  \dodoi{https://doi.org/10.1002/2013JA019537}

\bibitem[{Wang {et~al.}(2002)Wang, Ye, Wang, Zhou, \& Wang}]{Wang:2002}
Wang, Y.~M., Ye, P.~Z., Wang, S., Zhou, G.~P., \& Wang, J.~X. 2002, Journal of
  Geophysical Research: Space Physics, 107, SSH 2,
  \dodoi{https://doi.org/10.1029/2002JA009244}

\end{thebibliography}
\bibliographystyle{aasjournal}

\end{document}